\documentclass[twocolumn,prb,amsmath,amssymb,floatfix,superscriptaddress,showpacs]{revtex4}
\usepackage{graphicx}
\usepackage{dcolumn}
\usepackage{bm}

\usepackage{times}
\begin{document}

\setlength{\topmargin}{0in}

\title{Coexistence and competation of local- and long-range polar orders in a ferroelectric relaxor }
\author{Guangyong Xu}
\affiliation{Condensed Matter Physics and Materials Science
Department, Brookhaven National Laboratory, Upton, New York 11973}
\author{P. M. Gehring}
\affiliation{NIST Center for Neutron Research, National Institute of
Standards and Technology, Gaithersburg, Maryland, 20899}
\author{G. Shirane}
\affiliation{Physics Department, Brookhaven National Laboratory,
Upton, New York 11973}
\date{\today}

\begin{abstract}
We have performed a series of neutron diffuse scattering
measurements on a single crystal of the solid solution
Pb(Zn$_{1/3}$Nb$_{2/3}$)O$_3$ (PZN) doped with 8\%\ PbTiO$_3$ (PT),
a relaxor compound with a Curie temperature T$_C \sim 450$~K, 
in an effort to study the change in local polar orders from the polar
nanoregions (PNR) when the material enters the ferroelectric phase.
The diffuse scattering intensity increases monotonically upon
cooling in zero field, while the rate of increase varies dramatically around 
different Bragg peaks.  These results can be explained by
assuming that corresponding changes occur in the ratio of the optic
and acoustic components of the atomic displacements within the PNR.
Cooling in the presence of a modest electric field $\vec{E}$
oriented along the [111] direction alters the shape of diffuse
scattering in reciprocal space, but does not eliminate the
scattering as would be expected in the case of a classic
ferroelectric material.  This suggests that a field-induced
redistribution of the PNR has taken place.
\end{abstract}

\pacs{77.80.-e, 77.84.Dy, 61.12.Ex}

\maketitle

\section{Introduction}

Lead perovskite relaxor systems such as
Pb(Zn$_{1/3}$Nb$_{2/3}$)O$_3$ (PZN) and
Pb(Mg$_{1/3}$Nb$_{2/3}$)O$_3$ (PMN) have been the focus of intense
scientific scrutiny in recent years because they display
exceptionally strong dielectric and piezoelectric
properties~\cite{PZT1}.  These systems are marked by a strong,
frequency-dependent dielectric permittivity that exhibits relaxation
processes on many different time scales.  This relaxational
character has been attributed to the presence of small-scale regions
of the lattice that possess randomly oriented, local polarizations,
and which first appear at a temperature $T_d$, often referred to as
the Burns temperature~\cite{Burns}.  In PMN, for example, $T_d$
($\sim$650~K) lies a few hundred degrees above $T_m$, the temperature at
which the dielectric permittivity is largest. Neutron scattering
studies by Naberezhnov {\it et al.} have demonstrated the onset of
strong diffuse scattering in PMN near $T_d$, thus indicating a
strong link with the PNR.  These polar nanoregions, or PNR, are
believed to be several nanometers in size, grow with cooling, and
are widely viewed to play important roles in determining relaxor
properties~\cite{Cross}.  For this reason we have undertaken a
comprehensive study of the PNR properties using neutron diffuse
scattering techniques.

Diffuse scattering is very sensitive to local inhomogeneities and
short-range order in solid materials.  It is thus an extremely
powerful tool with which to study the structure of the PNR.  Neutron
and x-ray elastic diffuse scattering measurements on relaxor systems
can provide important information about the size, shape, and
polarization of the PNR.  The data collected by Vakhrushev {\it et
al.}~\cite{PMN_neutron3} on the diffuse scattering intensity
measured along directions transverse to the wavevector $\vec{Q}$
near many Bragg peaks in PMN provided the first quantitative study
of the relative magnitudes of the atomic displacements
(polarizations) within the PNR.  In a subsequent study by Hirota
{\it et al.}~\cite{PMN_diffuse} the authors proposed the concept of
the uniform phase shift, whereby the atomic displacements
responsible for the PNR could be decomposed into two components: an
acoustic term that corresponds to a uniform shift of the entire PNR
relative to the surrounding lattice, and an optic term that results
from the condensation of a transverse optic (TO) phonon.  Two other
important features of the PNR are the directions of the
polarizations and the shape of the PNR.  These issues have also been
extensively studied by diffuse scattering and other 
techniques~\cite{Xu_diffuse,Hiro_diffuse,PZN_diffuse3,PMN_diffuse2,
PMN_xraydiffuse,PMN_xraydiffuse2,Egame_PDF}.  Recently, Xu~{\it et
al.}~\cite{Xu_3D} studied the three-dimensional diffuse scattering
distribution in single crystals of PZN and its derivatives. These
authors found that the diffuse scattering consists of six $\langle
110\rangle$ rods.  A model was proposed where $\langle 1\bar{1}0
\rangle$-type polarizations are correlated in \{110\} planes, thus
implying a ``pancake'' shape for the PNR with polarizations that lie
in-plane.

Of course, with diffuse scattering measurements, it is not always easy
to distinguish subtle differences in the local structure. In other words,
are these PNR really nano-meter sized polar domains with well-defined 
boundaries; or just local polar fields with short-range correlations? The 
former correspond to a square-function type polar correlation, while the latter
can be described with a gradually decaying, e.g., exponentially decaying 
correlation function with a certain length scale. In both cases, local polar 
moments due to optic type atomic displacements exist in the system, as 
proposed by Burns and Dacol in their original PNR picture; and both can result 
in very similar diffuse scattering line shapes in thee reciprocal space. In this
paper, we do not attempt to make such a distinction. For simplicity, the term
``PNR'' is used throughout the paper. And when the ``size'' or ``shape'' of 
the PNR is concerned, they can be viewed alternatively as length scales 
describing the local polar field (instead of the size of a well defined 
nano-region).

Considered by many researchers to be precursors to the ferroelectric
phase transition, the PNR were expected to co-align or form much
larger ferroelectric domains when the system is driven into a
ferroelectric phase, either by cooling or by application of an
external dc electric field.  However diffuse scattering measurements
have provided little such indication.  Upon cooling in zero field,
the diffuse scattering intensities increase monotonically in both
PZN~\cite{Stock1} and PMN~\cite{Xu_diffuse} while the shape of the
scattering remains the same, suggesting that the PNR persist at 
low temperatures. The effects of an external dc field on the
diffuse scattering are more complicated; some
studies~\cite{PMN_efield,PZN-8PT} show a partial suppression of the
diffuse scattering measured along the [110] and [001] directions,
while more recent work~\cite{Xu_new, Xu_nm1} indicate a
redistribution of the PNR takes place when subjected to an external
field oriented along the [111]  direction. These studies indicate that
in relaxor systems, that local- and long-range polar orders  
coexist and compete with each other. 

The unit cells of pure PZN~\cite{PZN_Xu,PZN_Xu2} and
PMN~\cite{Bonneau,Husson} remain cubic when cooled in zero field;
thus neither compound exhibits long-range ferroelectric order at low
temperature.  However, doping with PbTiO$_3$ (PT) to form the solid
solutions PZN-$x$PT and PMN-$x$PT gradually suppresses the relaxor
character and establishes a ferroelectric phase~\cite{PMN_Ye}. By
increasing the PT content each system can be driven across a
morphotropic phase boundary
(MPB)~\cite{PZN_phase2,PMN_phase,PZN_phase} where a more
conventional ferroelectric phase~\cite{Feng,Ohwa} is achieved.  In
this paper, we report neutron diffuse scattering measurements on
PZN-8\%PT, which does undergo a ferroelectric phase transition upon zero-field 
cooling. The low temperature phase has a rhombohedral structure with 
$\langle111\rangle$ type polarizations. 
Being on the relaxor side of the phase diagram, PZN-8\%PT
also exhibits relaxor properties and strong diffuse
scattering~\cite{PZN-8PT,Xu_3D,Xu_new}.  The 8\%PT content is also
the composition at which the piezoelectric response is
maximum~\cite{PZN_phase2}, thus making the PZN-8\%PT compound a
particularly interesting choice of system in which to study the PNR
in a ferroelectric phase.

We have studied how the diffuse scattering changes when PZN-8\%PT is
cooled into the ferroelectric phase under (i) zero field and (ii) an
external field oriented along the [111] direction.  The
zero-field-cooling (ZFC) data show that the diffuse scattering has
the same shape in the high temperature paraelectric and low
temperature ferroelectric phases.  The observed increase in total
diffuse scattering intensity on cooling indicates a corresponding
growth in the total PNR volume, or an increase in the PNR
polarization, or both. However, there is a change in the ratio of
the acoustic and optic components of the atomic displacements in the
PNR. Specifically, the uniform shift of the PNR grows with
decreasing temperature, which suggests that a ``pinning'' effect
takes place in the low temperature phase.  After field cooling (FC),
a clear redistribution of the diffuse scattering intensity is
observed around all Bragg peaks, while the total (integrated)
diffuse scattering intensity appears to be conserved, i.\ e.\ is at
least equal to that in the ZFC case.

\section{Experiment}

The PZN-8\%PT single crystal used in this study is rectangular in
shape, having dimensions of $5 \times 5 \times 3$~mm$^3$ with {111},
{$\bar{2}$11}, and {0$\bar{1}$1} surfaces, and was provided by TRS
Ceramics.  Cr/Au electrodes were sputtered onto the top and bottom
{111} crystal surfaces.  This is the same crystal that was used in
Ref.~\onlinecite{Xu_new}, which reported structural transitions from
a cubic to tetragonal phase at $T_C \approx 440~$K, and then to a
rhombohedral phase at $T_{C2} \approx 340$~K in zero field.  Upon
cooling in a moderate electric field of $E=2$~kV/cm oriented along
the [111] direction, the value of $T_C \approx 460$~K, while $T_{C2}
\approx 340$~K does not change.

The neutron diffuse scattering measurements were performed with the
BT9 triple-axis spectrometer located at the NIST Center for Neutron
Research.  The measurements were made using a fixed incident neutron
energy $E_i$ of 14.7~meV, obtained from the (002) Bragg reflection
of a highly-oriented pyrolytic graphic (HOPG) monochromator, and
horizontal beam collimations of 40'-40'-S-40'-80' (''S'' = sample).
The (002) reflection of an HOPG analyzer was used to select the
energy of the scattered neutron beam.  Two PG filters were placed
before and after the sample to reduce the scattering from higher
order reflections.  Data were taken in the (HKK) scattering plane,
defined by the two primary vectors [100] and [011], with the [111]
electric field direction lying in the plane and the {0$\bar{1}$1}
crystal surface pointing vertically. All measurements were performed
while cooling, starting from 550~K so that all residual (poling)
effects are removed.

\section{Results and Discussion}

\begin{figure}[ht]
\includegraphics[width=\linewidth]{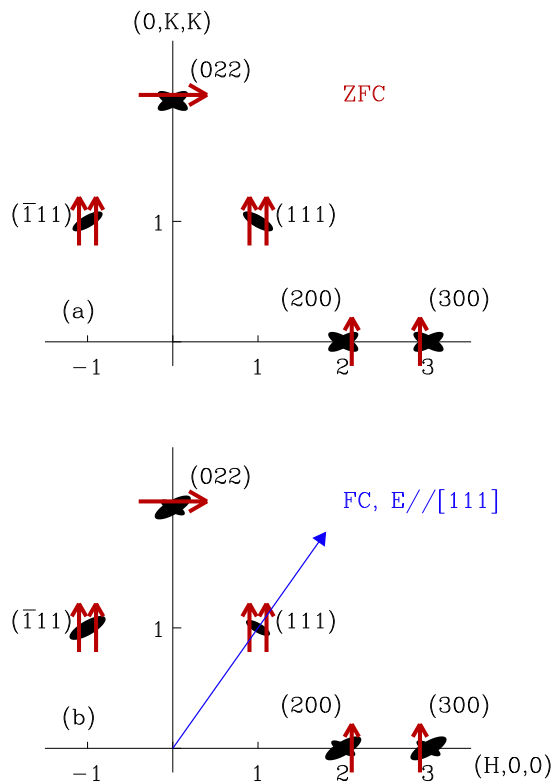}
\caption{(Color online) A schematic of the diffuse scattering
intensity distribution in PZN-8\%PT measured in the (HKK) scattering
plane around the (200), (300), (111), ($\bar{1}$11), and (022) Bragg
peaks at 300 K after (a) zero-field cooling, and (b) field cooling
with E=2~kV/cm along [111]. The arrows indicate the linear scans
performed in our measurements.} \label{fig:1}
\end{figure}

Previous measurements~\cite{Xu_new} have confirmed that the diffuse
scattering intensities in  PZN-8\%PT form ``butterfly'' shaped
patterns in zero field in the (HKK) scattering plane; this is shown
schematically in Fig.~\ref{fig:1}~(a).  These butterfly patterns are
in fact the result of the tails of out-of-plane $\langle 110\rangle$
rods of diffuse scattering intensity, which are visible because of
the non-zero instrumental out-of-plane wavevector ($q$) resolution.
The ``$\slash$''-shaped wing of the butterfly pattern measures the
tails of diffuse rods oriented along the [110] and [101] directions,
which arise from PNR having polarizations pointing along the
[1$\bar{1}$0] and [10$\bar{1}$] directions; the
``$\backslash$''-shaped wing measures the tails of diffuse rods
oriented along the [1$\bar{1}$0] and [10$\bar{1}$] directions, which
arise from PNR having polarizations pointing along the [110] and
[101] directions.  An efficient way to monitor changes in the
diffuse scattering is to perform linear $q$-scans in reciprocal
space that are offset from the Bragg peak, as shown by the (red)
arrows in Fig.~\ref{fig:1}.  These linear scans have the advantage
of being able to monitor the intensity changes in both the
``$\slash$'' and ``$\backslash$'' wings of the butterfly pattern,
which would not be possible with a (transverse) linear scan that cut
across the Bragg peak itself.

\begin{figure}[ht]
\includegraphics[width=\linewidth]{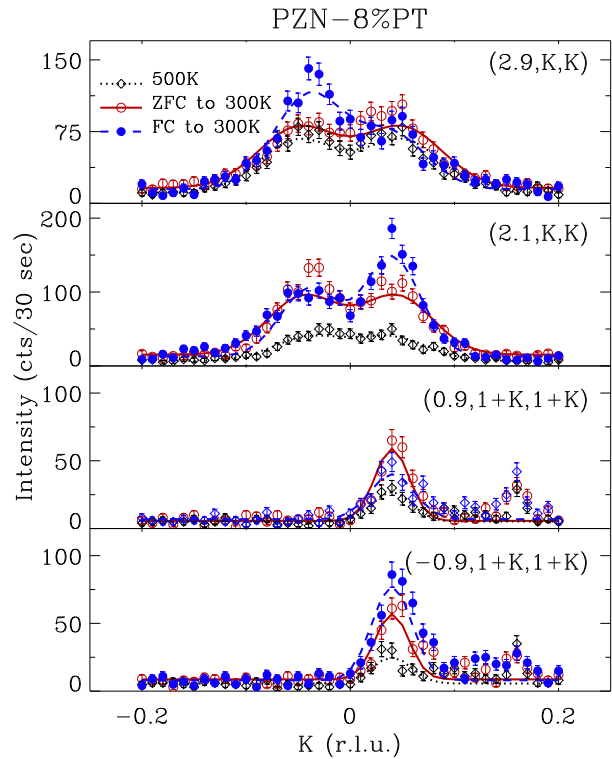}
\caption{(Color online) Diffuse intensity linear profiles, measured
along the [011] direction around the (300), (200), (111), and
($\bar{1}$11) peaks, are shown from top to bottom respectively. The
lines through the data are fits to Gaussian functions of the reduced
wavevector $q$ and are merely guides to the eye. The weak peaks in the (111)
and ($\bar{1}$11) scans near $K\sim0.17$ are from aluminum powder lines. } 
\label{fig:2}
\end{figure}

Some of these linear scans are plotted in Fig.~\ref{fig:2}.  We note
that the linear scans near (300) (measured along (2.9,K,K)) and
(200) (measured along (2.1,K,K)) produce a double-peak profile
because the diffuse scattering intensity peaks at each wing of the
butterfly pattern.  By contrast the scans near (111) (measured along
(0.9,K,K)) and ($\bar{1}$11) (measured along (-0.9,K,K)) only have
one peak. This agrees well with the polarization analysis; the
``$\slash$'' wing comes from those PNR with polarizations
perpendicular to [111], and thus is not observed around the (111)
Bragg peak; neither is the $\backslash$ wing observed around the
($\bar{1}$11) peak.

In the following subsection, we study the change of the diffuse
scattering intensity profiles when the system is driven into a
ferroelectric phase by ZFC (section III A) and FC (section III B).

\subsection{Temperature Effects}

Upon ZFC from 500~K to 300~K, i.\ e.\ below $T_c$ the diffuse
scattering intensities measured around all Bragg peaks increases.
However, the rates of increase are very different.  The diffuse
scattering intensities around the (300) peak increase slightly from
500~K to 300~K, while those around the (200) peak grow by almost a
factor of two. Similar behavior was also observed in the relaxor
system PMN, where the diffuse scattering intensities around (200)
peak increase much faster with cooling than do those measured around
the (300) peak~\cite{Hiro_private}. In order to understand this
change of relative diffuse scattering intensities across different
Bragg peaks, we first consider the various factors that determine
the diffuse scattering intensities observed in PZN-8\%PT.

Diffuse scattering refers to the relatively weak intensity that
decorates broad regions of reciprocal space around different Bragg
peaks.  One source of this scattering arises from short-range
correlated displacements of atoms from the ideal lattice positions.
The diffuse scattering intensity distribution due to the PNR in
PZN-8\%PT measured near the Bragg vector ${\bf G}$ at the wavevector
${\bf Q} = {\bf G} +{\bf q}$ can be written as

\begin{equation}
I_{diff}({\bf Q})=A\sum_i|F_{diff}^i({\bf G})|^2|f_i({\bf q})|^2.
\label{eqn:1}
\end{equation}

Here the sum is taken over the contributions from all six possible
PNR orientations, which correspond to ``pancakes'' in real space
with six different \{110\} surfaces~\cite{Xu_3D}.  The term
$|f_i({\bf q})|^2$ describes the Fourier transform of the shape of
the $i$th orientation ($i = 1$~to~$6$) of the PNR, and depends only 
on ${\bf q}$.
This term gives the diffuse scattering a characteristic rod-like
shape in reciprocal space.  A single pair of $\langle 110
\rangle$-oriented rods produces the butterfly shaped diffuse
scattering contours around certain Bragg peaks.  On the other hand,
the diffuse scattering structure factor $|F_{diff}^i({\bf G})|^2$
gives the relative intensity of the diffuse scattering rod near the
Bragg peak represented by reciprocal lattice vector ${\bf G}$. This
quantity can be expressed in terms of the relative magnitudes of the
atomic displacements in the unit cell (within the $i$th orientation
of the PNR)

\begin{equation}
|F_{diff}^i({\bf G})|^2 =
\sum_k |{\bf Q}\cdot{\bf \xi_k^i}|^2 b_k\exp{(-W_K)}\exp{(i{\bf G}
\cdot{\bf R_k})},
\label{eqn:2}
\end{equation}

where ${\bf \xi_k^i} $ is the atomic displacement vector, $b_k$ is
the neutron scattering length of atom $k$, and $R_k$ is the lattice
position of the $k$th atom in the unit cell.  The Debye-Waller
factors $\exp{(-W_K)}$ vary relatively little over the temperature
range of our study and are thus neglected in the analysis presented
here. Here we take the model proposed in Ref.~\onlinecite{Xu_3D} that 
in-plane $\langle 1\bar{1}0\rangle$ atomic
displacements can result in the rod-type diffuse scattering along 
\{110\} directions.  We can therefore simplify the expression for the
structure factor as

\begin{equation}
|F_{diff}^i({\bf G})|^2 \propto |{\bf Q}\cdot{\bf \epsilon_i}|^2
\sum_k  b_k\xi_k\exp{(i{\bf G}\cdot{\bf R_k})},
\label{eqn:3}
\end{equation}

where ${\bf \epsilon_i}$ is the unit vector along the polarization
direction ($\langle 1\bar{1}0\rangle$) of the $i$th orientation of
the PNR.

\begin{table}
\caption{Integrated diffuse scattering intensity measured around the
(200), (300), and (111) Bragg peaks.}
\begin{ruledtabular}
\begin{tabular}{cccc}
    &   (200)   &   (300)   &   (111)   \\
\hline
$|{\bf Q}\cdot\hat{\epsilon}|^2$ & 2.0 & 4.5 & 2.0 \\
&&&\\
500 K   &   6.52    &   11.6    &   2.52\\
400 K   &   10.94   &   13.2    &   3.75\\
300 K   &   16.86   &   13.2    &   5.04 \\
\end{tabular}
\end{ruledtabular}
\label{tab:1}
\end{table}

Our measurements show that the diffuse scattering in PZN-8\%PT does
not change shape significantly upon ZFC below $T_C$, i.e. the
structure factor $|f_i({\bf q})|^2$ does not change qualitatively.
The diffuse scattering distribution still consists of six $\langle
110\rangle$-oriented rods, the widths and lengths of which may vary,
reflecting changes in the sizes or magnitude of polarizations (or both) of the
PNR, but these changes do not contribute to a change in the relative
intensities of the diffuse scattering measured near different Bragg
peaks.  If measured at the same offset ${\bf q}$ from a given Bragg
peak ${\bf G}$, the relative intensities of the diffuse scattering
should be completely determined by the term $|F_{diff}^i({\bf
G})|^2$.  The linear profiles shown in Figs.~\ref{fig:1} and
\ref{fig:2} were obtained at the same ${\bf q}$, thereby measuring
the intensities of the ``$\slash$'' and ``$\backslash$'' wings at
the same offset from the respective Bragg peaks. We can therefore
compare these intensities directly.  The $q$-integrated intensities
of the linear profiles are shown in Table~\ref{tab:1}. The
$q$-integrated intensities of linear profiles around (200) and (300)
include intensity contributions from both the ``$\slash$'' and
``$\backslash$'' wings, both of which have the same $|{\bf
Q}\cdot\hat{\epsilon}|^2$ factor.  At the (111) peak (and
($\bar{1}$11) peak), one of the wings is absent. In order to make a
direct comparison, we listed the sum of the integrated intensities
measured at (111) and ($\bar{1}$11) in the last column to include
the intensity of both wings.

\begin{table}
\caption{Calculated atomic displacements and relative intensities of
the different modes. The numbers for the Slater mode and Last mode,
and the uniform shift, correspond to the size of the oxygen
displacements.}
\begin{ruledtabular}
\begin{tabular}{ccccccc}
    &$\delta_{Pb}$&$\delta_{Zn,Nb}$ &$\delta_{O}$&Slater&Last&Shift\\
\hline
&&&\\
500 K   &1.0&0.30&-0.11&-0.24&-0.56&0.69\\
400 K   &1.0&0.37&-0.07&-0.26&-0.52&0.71\\
300 K   &1.0&0.46&-0.02&-0.28&-0.47&0.73\\
\end{tabular}
\end{ruledtabular}
\label{tab:2}
\end{table}

The relative magnitudes of the atomic displacements can therefore be
calculated from Eq.~\ref{eqn:3}. The results are shown in the first
three columns of Table~\ref{tab:2}. All numbers are normalized to
the Pb shifts.  Note that these numbers are relative, i.e., we do
not know if the increase of the overall diffuse scattering intensity
upon cooling is due to increasingly larger atomic displacements, a
larger total PNR volume, or both.  These results only provide
information on how the relative displacements of the Pb, Zn(Nb), and
O atoms change.  However, it is clear that upon cooling both the
Zn(Nb) and O atoms tend to shift in accordance with the Pb atom
shifts. This indicates a more ``acoustic'' type shift for the PNR.

Since the PNR result from the condensation of a soft transverse
optic (TO) phonon, the atomic displacements of the PNR contributing
to the diffuse scattering should be consistent with the normal mode
vibrational displacements associated with the TO phonon.  In the
lead perovskite relaxors, the dominant vibrational modes are the
Slater mode and Last mode~\cite{Harada}. In the Slater mode, the A
site atoms (Pb) remain stationary while the B site atoms (Zn/Nb)
move in opposition to the oxygen octahedra. In the Last mode, the B
site atoms and oxygen octahedra move together in opposition to the A
site atoms. However, in contrast to these two modes, the atomic
displacements listed in Table~\ref{tab:2} clearly violate the
condition that the unit cell center-of-mass be preserved.  Here we
use the same type of analysis as was carried out in
Ref.~\onlinecite{PMN_diffuse}, which included an additional acoustic
component, aka the ``Uniform Phase Shift.''  The relative ratios of
all modes can then be determined (see the last three columns in
Table~\ref{tab:2}). It is interesting to note that the ``Uniform
Phase Shift'' already exists at high temperatures in the
paraelectric phase.  Upon ZFC into the ferroelectric phase, this
uniform shift increases faster than does the growth of the
polarization (optic components) of the PNR.  We do not yet
understand the reason for this. Nevertheless, we can speculate that
in the low temperature ferroelectric phase, the PNR must be
uniformly shifted even further from the surrounding ferroelectric
polar environment in order to account for the increasing
electrostatic energy induced by the surrounding polar field. 

\subsection{Field Effects}

\begin{figure}[ht]
\includegraphics[width=\linewidth]{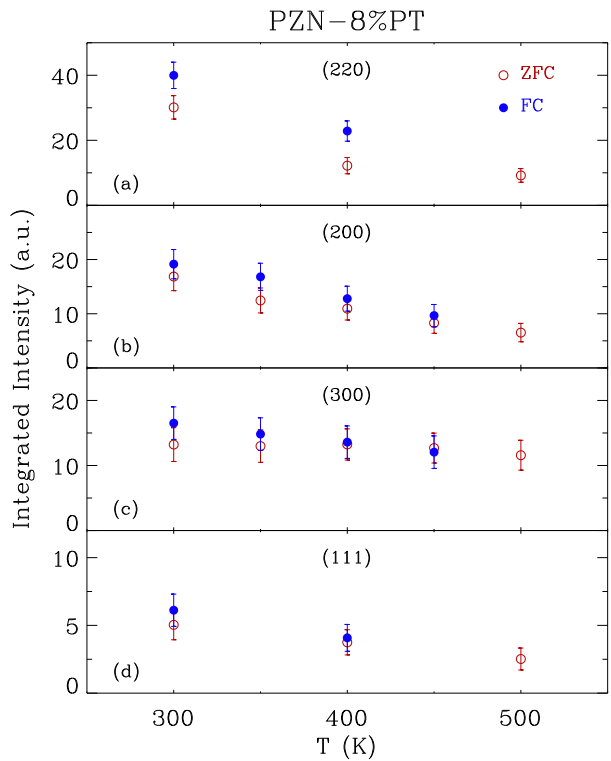}
\caption{(Color online) Integrated diffuse scattering intensity vs.
temperature measured near the (a) (220), (b) (200), (c) (300), and
(d) (111) Bragg peaks. The red open circles represent ZFC data while
the blue solid circles are FC data for E=2~kV/cm oriented along the
[111] direction. } \label{fig:3}
\end{figure}

In contrast to the ZFC case, cooling in the presence of a moderate
electric field ($E=2$~kV/cm) oriented along the [111] direction
changes the shape of the diffuse scattering dramatically (see
Fig.~\ref{fig:2}). All of the linear profiles show a common feature
whereby the ``$\slash$'' wing is enhanced and the ``$\backslash$''
wing suppressed. The overall change is plotted schematically in
Fig.~\ref{fig:1}~(b). This partial enhancement of the diffuse
scattering intensity was first reported in Ref.~\onlinecite{Xu_new}
and confirmed later by room temperature electric field measurements
on single crystals of pure PZN~\cite{Xu_nm1}. In this subsection we
analyze the quantitative change of the diffuse scattering induced by
field cooling.  The integrated intensities of the linear profiles
are plotted in Fig.~\ref{fig:3}. Here the intensities of both wings
are included (the numbers around the (111) peak are actually the
sums of the ``$\backslash$'' wing at the (111) peak and the
$\slash$'' wing at the ($\bar{1}$11) peak, as was done for
Table~\ref{tab:1}). We find that the $q$-integrated diffuse
intensities of both wings increase monotonically with cooling,
whether ZFC or FC.  The difference between the integrated
intensities in the ZFC and FC cases are relatively small,
considering the size of the error bars. This demonstrates that there
is very likely a redistribution of the diffuse scattering intensity
among the different $\langle 110\rangle$-oriented diffuse rods,
which must be associated with a redistribution of the PNR with
different orientations/polarizations.

A slight enhancement of the FC linear integrated intensity compared
to the ZFC case is also observed, as shown in Fig.~\ref{fig:3}.
Considering previous reports of a field-induced suppression of the
diffuse scattering intensities measured transverse to the Bragg
wavevector in both PMN and PZN-8\%PT~\cite{PMN_efield,PZN-8PT}, it
is possible that the shape of each individual $\langle 110\rangle$
diffuse rod is also slightly affected by the field. Therefore
measurements at different distances away from the Bragg peaks may
lead to slightly different results.  However, in general we find
that the diffuse scattering in PZN-8\%PT still consists of $\langle
110\rangle$ rods upon FC, and that the intensities are mostly
redistributed between different $\langle 110\rangle$ rods, instead
of within an individual rod.

\begin{figure}[ht]
\includegraphics[width=\linewidth]{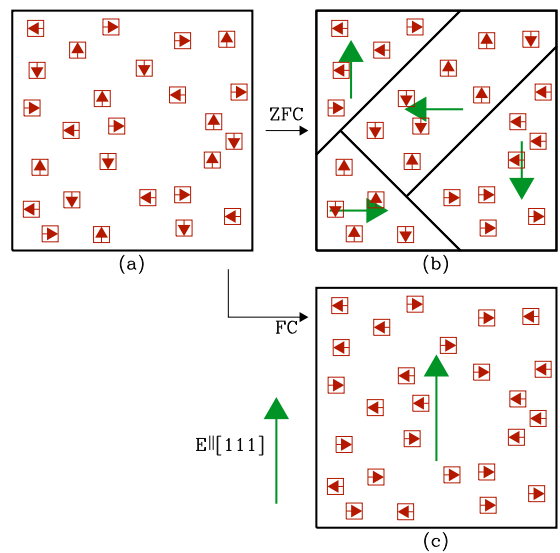}
\caption{(Color online) A schematic showing the PNR configurations
in a relaxor system in (a) the paraelectric phase, (b) ZFC into the
ferroelectric phase, and (c) FC into the ferroelectric phase. The
large arrows indicate the polarization of the ferroelectric domains
separated by domain walls (solid lines). The small squares represent
the PNR.} \label{fig:4}
\end{figure}

On the other hand, it is quite intriguing to note that the enhanced
``$\slash$'' wing comes from those PNR having polarizations (along
[1$\bar{1}$0] and [10$\bar{1}$]) perpendicular to the [111] field
direction. This goes against the natural expectation that the field
should be able to ``align'' the PNR to point along the field
direction. The same behavior was also observed in a room temperature
``poling'' experiment performed on a single crystal of PZN using a
[111] field~\cite{Xu_nm1}. Evidently, the energy of the PNR alone in
the electric field does not favor such a configuration.  This
suggests that the interaction between the PNR and the surrounding
lattice must also be taken into account.  In the paraelectric state
the PNR are randomly oriented along one of six $\langle110\rangle$
directions (see Fig.~\ref{fig:4} (a)).  
Below the ferroelectric phase transition, 
macroscopic ferroelectric
domains form and the cubic symmetry is broken. Thus PNR polarized
along the six $\langle110\rangle$ directions are no longer
equivalent. After cooling in zero field the polarizations of the
macroscopic ferroelectric domains are randomly distributed along any of the four
$\langle111\rangle$ directions such that there is no macroscopic
preferred $\langle110\rangle$ polarization of the PNR, as described
schematically in Fig.~\ref{fig:4} (b).  However,
cooling in field with E applied along [111] greatly enhances the
volume of the [111] polarized ferroelectric domain (Fig.~\ref{fig:4} (c)). 
Our current results are consistent with the room temperature [111] field poling
measurements on PZN.  Both exhibit an alignment of the ferroelectric
domains, instead of the PNR themselves, due to the external electric
field. Our data show that those PNR with polarizations perpendicular
to the field, and therefore perpendicular to the polarization of the
domain in which they reside, are enhanced. This may in fact imply
that these PNR are the preferred configuration in the ferroelectric
phase transition.  The application of a [111] field rearranges
ferroelectric domains, which in turn reveals this underlying bias of
the PNR macroscopically.

\section{Conclusions}

Our diffuse scattering measurements on PZN-8\%PT suggest that PNR
persist into the low temperature ferroelectric phase under both ZFC
and FC conditions.  This is consistent with previous diffuse
scattering measurements as well as Raman
results~\cite{Toulouse_KLT,Toulouse_KLT2,Toulouse_KLT3} obtained on
other relaxor systems.  We  show that local atomic displacements
contributing to the diffuse scattering consist of 
both an acoustic (strain) and optic (polar) component. This provides 
convincing evidence of the ``polar'' nature of these local orders, and
stands against
the argument that diffuse scatterings in relaxor systems are mainly 
due to local strain fields instead of local polarizations. 
In addition, the acoustic component, or the  ``Uniform Phase Shift'' of 
the PNR increases faster than do the polarizations upon ZFC.  The PNR are
therefore more strongly displaced in the ferroelectric phase,
causing the structure of the PNR to be more ``out-of-phase'' from
the surrounding environment.  

Our FC results reveal that the
preferred configuration of the PNR in the ferroelectric phase is
such that the PNR tend to align perpendicular to the polarization of
the surrounding ferroelectric polar domain. This unusual
configuration makes the polarization of PNR ``out-of-phase''
from the surrounding environment.  These are probably the two most
important factors that help to keep preserve the coexistence of 
PNR and the long range polar order in the ferroelectric phase.
On the other hand, with both short-range polar order in the PNR and 
long-range ferroelectric order in the lattice developing at the same 
time (upon cooling), the strain caused by this mismatch will increase, 
resulting in an increase of the local strain field, which corresponds to the 
acoustic component of the local atomic displacements.

The coexistence of the short-range polar order of the PNR and the
long range ferroelectric order of the ferroelectric phase is quite
subtle. It can be affected by many factors such as electric fields -
where we believe a strong enough field should be able to eventually
directly affect the PNR; pressure - which appears to be able to
suppress the diffuse scattering from PMN~\cite{Pressure} and
PZN~\cite{Pressure_PZN}; and doping with PT - which eventually
changes the structure of the low temperature phase and inevitably
affects the PNR. These are interesting future topics that will
require more detailed quantitative study, and which will be
challenging and important.

\begin{acknowledgements}
We would like to thank H.~Hiraka and C.~Stock, for stimulating
discussions. Financial support from the U.S. Department of Energy
under contract No.~DE-AC02-98CH10886 is also gratefully
acknowledged.
\end{acknowledgements}


\end{document}